# Ultrafast charge dynamics and photoluminescence in bilayer MoS$_2$

Naseem Ud Din, Volodymyr Turkowski, and Talat S. Rahman

Department of Physics, University of Central Florida, Orlando, FL 32816.

Our examination of the interplay of ultrafast charge dynamics and electron-phonon interaction in bilayer MoS$_2$ provides a microscopic basis for understanding the features (two peaks) in the emission spectrum. We demonstrate that while the initial accumulation of excited charge occurs at and near the Q point of the two-dimensional Brillioun zone, emission takes place predominantly through two pathways: direct charge recombination at the K point and indirect phonon-assisted recombination of electrons at the K valley and holes at Γ hill of the Brillouin zone. Analysis of the wave vector dependencies of the electron-phonon interaction traces the higher energy peak to phonon-assisted relaxation of the excited electrons from the Q to the K valley in the conduction band. Our results thus reveal the importance of ultrafast charge dynamics in understanding photoemissive properties of a few-layer transition-metal dichalcogenide. These calculations are based on time dependent density functional theory in the density matrix formulation.

1. **Introduction**

Transition metal dichalcogenides (TMDCs) form a layered structure with strong in-plane and weak out-of-plane interaction [1, 2]. As is now well known, in TMDCs such as MoS$_2$, MoSe$_2$, WS$_2$ and WSe$_2$ the bandgap that changes from indirect to direct as the thickness of the TMDC reduces to that of a single layer, making it suitable for applications in transistors, photodetectors and electroluminescent devices [2-9]. Single-layers of the above mentioned TMDCs, in particular of MoS$_2$, have received significant attention also because of their extremely high photoluminescence (PL), a band gap that lies in the visible spectrum, and strong binding energy of excitons [3, 4].

Bilayer molybdenum disulfide is also a system with remarkable characteristics. It has even higher electron mobility and density of states, as compared to single-layer, showing great potential for DC and high-frequency electronic applications [10], and with Bernal stacking, sensitive to electric field. Bilayer MoS$_2$ also demonstrates strong PL [4, 11-14], and similar to the monolayer, it has strongly bound (~100-300 meV) excitonic states [15-18]. However, unlike the single layer, bilayer MoS$_2$ displays two distinct peaks in the PL: one at an energy below (~1.6 eV) and the other at the same energy (~ 1.9 eV) as that for the single layer [4, 16]. While the peak around 1.9 eV, emerging from the direct band gap in single layer MoS$_2$, remains unchanged for the bilayer, the dependence of the lower energy peak, corresponding to the indirect band gap, has been the subject of several investigations [16, 19] in which the nature of stacking and twist angle between the two layers has been varied [16]. These studies confirm the existence of weak interlayer tunnel coupling and strong intralayer electrostatic coupling [19], typical of the TMDCs. They also point to tunability of the interlayer coupling by twisting the layers relative to each other, providing one more avenue for manipulating the optoelectronic properties of few layer MoS$_2$. Transport [20], emissivity[21], and electronic [22] properties may be tuned by pressure, and Berry curvature, and spin and valley Hall effects, may be manipulated by an applied electric field [23]. With promising optoelectronic applications, the characteristics of the charge carriers and their bound forms, excitons, biexcitons and trions have been examined [24-27]. The role of characteristic vibrational modes [28-30] and electron-phonon coupling [30] in controlling the optical properties on few layer TMDCs has also been the subject of intense research and debate (see also [18] and references therein).

More germane to this work, analysis of properties at the picosecond timescale of a few-layer MoS$_2$, showed [31] that nonradiative relaxation mechanisms dominate the carrier dynamics in the system, leading to the speculation that dynamical processes such as valley charge redistribution will come into play before radiative emission. In fact, application of bias voltage was found to drive charges between different band minima [11], resulting in sub-room temperature dissociation of trions which, for the single-layer are stable at room temperature. Although the timescale of involved dynamical processes was not discussed by Kümmell et al. [11], the above results suggest that bias voltage can be a knob for tuning exciton and trion emission through transfer of charge between the layers. In related theoretical work a phenomenological model was proposed [32] to analyze the picosecond valley depolarization dynamics under an external electric-field. Tight binding models have also been applied to examine inter-subband transition rates in e- and h-doped systems [33].

Illustrative as the above experimental and theoretical studies have been in exposing the band structure, vibrational dynamics, and optical properties of 2L MoS$_2$, they have not tracked the microscopic processes responsible for the emissive properties of 2L MoS$_2$. In particular, the following questions have not been addressed: What is the effect of electron-phonon coupling on the intervalley charge dynamics at ultrafast time scales? What are the contributions of the different valleys to the emission spectrum? Answers to these questions would provide a systemic understanding of the response of 2L MoS$_2$ to an ultrafast (femtosecond) pulse, which would in turn help manipulate system properties for potential ultrafast applications. With the above in mind, we have applied time dependent density functional theory (TDDFT) in the density matrix formalism to examine the temporal evolution of excited charges in the presence of electron-phonon interaction and their impact on the emission spectrum. To isolate the effect of electron-phonon interactions, we have neglected electron-electron and electron-hole interactions (i.e. excitonic effects are ignored), whose inclusion would have small quantitative effect and will not change the main conclusions in this work. We calculate phonon dispersion curves and the electron-phonon coupling constant using density functional perturbation theory (DFPT), within the harmonic approximation. We then include electron-phonon interaction phenomenologically in the Liouville equation via additive many-body scattering terms, in the spirit of semiconductor Bloch equations [27]. The details of computational methods are provided in section 2. For completeness, the calculated electronic structure is summarized in section 3. Results for the electron-phonon coupling coefficients are discussed in section 4, and that for the relaxation of excitations are presented in section 5, while the calculated emission spectra are presented in section 6. Our conclusions are summarized in section 7.

## 2. Theoretical and Computational Methods

### 2.1 Calculations of Electronic Structure:

We performed calculations based on DFT with the plane-wave and pseudopotential methods as implemented in the Quantum Espresso package [34]. We treated exchange correlation effects within the generalized gradient approximation in the form of Perdew–Burke–Ernzerhof (GGA-PBE) [35], and used ultrasoft pseudopotentials to describe the core-valence interactions. As already mentioned, 2L MoS$_2$ is a material with van der Waals inter-layer interaction, whose contribution we included using the vdW-DF2 method [36]. To mimic single-layer and bilayer, we applied periodic boundary conditions along x and y-axis and added a 15 Å vacuum along z-axis to eliminate any interaction of the system with its periodic image. For calculations of the bulk system we used two MoS$_2$ formula units (12.37 Å thickness) and applied periodic boundary conditions along all three coordinate axes. Side and top view of the bilayer are presented schematically in Figure 1. We described the valence wave functions, and electron density by plane-wave basis sets with kinetic energy cutoffs of 60 Ry and 360 Ry, respectively. We sampled the first Brillouin zone with a 14 × 14 × 14 Monkhorst-Pack grid for the bulk system and 14 × 14 × 1 grid for the single and bilayer systems. We optimized atomic positions and lattice parameters, until the residual forces converged to less than 0.01 eV/Å.

Note that at the outset we include spin-orbit coupling (SOC) by treating the core electrons fully relativistically. These SOC calculations are essential for identifying the splitting at the Brillouin zone edge and at the top of the valence band for the single layer. We thus quantified the splitting (between the spin-up and spin–down bands) of valence band edge at the K point to be 150 meV, in good agreement with the experimental value of 141 meV [37], and verified that they display opposite spin ordering at K and K' points. For the remaining calculations we chose not to distinguish the spins of the excited electrons and thus did not include SOC in the results that are presented below. The effect of SOC on the observables of interest here is negligible, and not worth the extra computational cost.

## 2.2 Calculations of the phonon spectrum and electron-phonon coupling coefficients:

We used density functional perturbation theory (DFPT) as implemented in the Quantum Espresso code [34], to first calculate the dispersion of the phonons across the Brillioun zone. For calculations of phonon frequencies, the residual forces were converged to less than 0.0001 eV/Å. Next, we calculated the electron-phonon coupling coefficients using the DFT results for the ground-state atomic and electronic configurations, the corresponding wave functions and band structure, and the calculated phonon dispersion. The quantities of interest are the electron-phonon scattering coefficients given by

$$g_{\vec{q}v}(k, i, j) = \left(\frac{\hbar}{2M\omega_{\vec{q}v}}\right)^{1/2} \left\langle \psi^i_{\vec{k}} \left| \frac{dV_{SCF}}{d\hat{u}_{\vec{q}v}} \cdot \hat{\epsilon}_{\vec{q}v} \right| \psi^j_{\vec{k}+\vec{q}} \right\rangle \qquad (1)$$

which correspond to the scattering of electron from state i (momentum $\vec{k}$) to state j (momentum $\vec{k} + \vec{q}$) due to absorption (emission) of phonon with mode index $v$ and momentum $\vec{q}(-\vec{q})$. In Eq. (1) M is atomic mass; $\psi^i_{\vec{k}}$ and $\psi^j_{\vec{k}+\vec{q}}$ are the electronic wavefunctions for the initial and final states, respectively; $\frac{dV_{SCF}}{d\hat{u}_{\vec{q}v}}$ is the gradient of the self-consistent potential with respect to the atomic displacements induced by the phonon mode $(\vec{q},v)$ with frequency $\omega_{\vec{q}v}$ and polarization vector $\hat{\epsilon}_{\vec{q}v}$. With the above definition of $g_{\vec{q}v}(\vec{k}, i, j)$, one can obtain the phonon line widths:

$$\gamma_{\vec{q}v} = 2\pi\omega_{\vec{k}v}\sum_{i,j}\int\frac{d^3k}{\Omega_{BZ}}|g_{\vec{q}v}(\vec{k},i,j)|^2\delta(\varepsilon_{\vec{q}}^i - \varepsilon_F)\delta(\varepsilon_{\vec{k}+\vec{q}}^j - \varepsilon_F), \quad (2)$$

where $\Omega_{BZ}$ is the volume of the first BZ, $\varepsilon_{\vec{q}}^i$ is the energy of the electron in the state (band) i and with momentum $\vec{q}$, and $\varepsilon_F$ is the Fermi energy. This brings us to the electron-phonon coupling constant for the corresponding phonon mode v with the wave vector q:

$$\lambda_{\vec{q}v} = \frac{\gamma_{\vec{q}v}}{\pi\hbar N(\varepsilon_F)\omega_{\vec{q}v}^2} \quad (3)$$

where $N(\varepsilon_F)$ is the electron DOS at the Fermi level. Using the results for the coupling constant (3) and phonon dispersion $\omega_{qv}$, one can obtain the isotropic Eliashberg spectral function:

$$\alpha^2F(\omega) = \frac{1}{2}\sum_v\int\frac{d^3q}{\Omega_{BZ}}\omega_{\vec{q}v}\lambda_{\vec{q}v}\delta(\omega - \omega_{\vec{q}v}). \quad (4)$$

**2.3 Calculation of time dependent excited state charge densities the emission spectra:**

Input from above DFT calculations form the basis for the code based on Density-Matrix Time Dependent Density Functional Theory (DM -TDDFT) [38, 39] that we used to calculate excited state charge densities and the emission spectrum. However, as mentioned above the exchange-correlation kernel was set to zero. Details of the calculations of the time-dependence of the excited charge density may be found in the Supplementary Information (SI) and in Ref [40] (in which the formalism is given for the more general spin-polarized case).

To study the effect of electron-phonon coupling on excited-charge dynamics we solved the density-matrix Liouville equation with the scattering term (Bloch equations for semiconductors):

$$\frac{\partial\rho_k^{lm}(t)}{\partial t} = [H,\rho]_k^{lm}(t) \equiv \sum_n\left(H_k^{ln}(t)\rho_k^{nm}(t) - \rho_k^{ln}(t)H_k^{nm}(t)\right) + \left(\frac{\partial\rho_k^{lm}}{\partial t}\right)_{scatt}, \quad (5)$$

where

$$H_k^{ml}(t) = \int\psi_k^{m*}(r)\hat{H}(r,t)\psi_k^l(r)dr \quad (6)$$

are the matrix elements of the Kohn-Sham Hamiltonian with respect to the static DFT wave functions, and $\left(\frac{\partial \rho_{\vec{k}}^{lm}}{\partial t}\right)_{scatt}$ are the scattering terms whose details are provided in the SI. Recall that the inherent relationship between the wavefunction and density matrices follows from the time dependent Schrodinger equation $i\frac{\partial \Psi_{\vec{k}}(\vec{r},t)}{\partial t} = H(\vec{r},t)\Psi_{\vec{k}}(\vec{r},t)$ when one uses the ansatz $\Psi_{\vec{k}}(\vec{r},t) = \sum_l c_{\vec{k}}^l(t)\psi_{\vec{k}}^l(\vec{r})$, where $\psi_{\vec{k}}^l(\vec{r})$ and $c_{\vec{k}}^l(t)$ are the static DFT wavefunctions and time-dependent coefficients, respectively (l is the band index, k is the wave-vector). the problem then reduces to finding the coefficients $c_{\vec{k}}^l(t)$, or their bilinear combination $\rho_{\vec{k}}^{lm}(t) = c_{\vec{k}}^l(t)c_{\vec{k}}^{m*}(t)$, the density matrices, which satisfy the Liouville equation.

The strength of the transition dipoles corresponding to the photon-induced electronic transitions was calculated from the dipole moment $\vec{d}_{\vec{k}}^{lm}$:

$$\left|\vec{d}_{\vec{k}}^{lm}\right|^2 = \left|e\int \psi_{\vec{k}}^{l*}(\vec{r})\vec{r}\psi_{\vec{k}}^m(\vec{r})dr\right|^2. \tag{7}$$

The electron-phonon coupling terms defined in Eqs. (1) - (3) were used to calculate the scattering terms in the TDDFT Bloch equations from the many-body Bloch equations derived for the electron-phonon part of the Hamiltonian $H_{e-ph} = \sum_{l,\vec{k},\vec{q},\nu} \hbar g_{\vec{q}\nu} a_{\vec{k}+\vec{q}}^{l+} a_{\vec{k}}^l (b_{\vec{q}}^\nu + b_{-\vec{q}}^{\nu+})$ (a and b are the fermion and phonon operators, correspondingly, l and $\nu$ are the electron and the phonon band indices and $\vec{k},\vec{q}$ are momenta. For more details, we refer the reader to the SI and to Ref. [27].

We next calculated the absorption spectrum: $A(\omega) = \frac{\omega}{n_b c} Im[\epsilon_{mac}(\omega)]$, where c is speed of light, $n_b$ is the background refractive index, and

$$\epsilon_{mac}(\omega) = 1 - 4\pi\chi_{KS}(\vec{k} \to 0, \omega), \tag{8}$$

is the macroscopic dielectric function related to the Kohn-Sham DFT susceptibility $\chi_{KS}$ as below:

$$\chi_{KS}(\vec{k},\omega) = \frac{1}{V}\sum_{\vec{k}'}\sum_{j=1}^{\infty}\sum_{l=1}^{\infty} \frac{\left(f_{\vec{k}+\vec{k}'}^l - f_{\vec{k}'}^j\right)}{\omega + \varepsilon_{\vec{k}'}^j - \varepsilon_{\vec{k}+\vec{k}'}^l + i\delta} \int d^3r \psi_{\vec{k}'}^{j*}(\vec{r})\vec{r}\psi_{\vec{k}+\vec{k}'}^l(\vec{r})$$
$$\times \int d^3r' \psi_{\vec{k}+\vec{k}'}^{l*}(\vec{r}')\vec{r}'\psi_{\vec{k}+\vec{k}'}^j(\vec{r}'). \tag{9}$$

In Eq. (9), $\varepsilon_{\vec{k}}^{l}$ and $\psi_{\vec{k}}^{l}(\vec{r})$ are the Kohn-Sham DFT eigenenergies and eigenfunctions, $f_{\vec{k}}^{l}$ is the Fermi factor for the corresponding DFT state; and j and l are the (valence and conduction) band indices. In the results below we have included two valence and two conduction bands in the summations in Eq. (9). Further details may be found in Ref [41].

The emission spectrum is then obtained by simply multiplying the absorption spectrum A(ω) by the Planck factor [42]:

$$E(\omega) = \frac{4\pi\omega^4}{e^{\frac{\omega-\Delta}{T}}-1} A(\omega), \tag{10}$$

where Δ is the optical gap in the system and the temperature is taken to be $T = 0.01 eV$, a value of order of room temperature. Furthermore, we have incorporated the nonequilibrium excited charge density in the calculation of the emission spectrum. This is accomplished by multiplying the expression on the right-hand side of Eq. (10) by the total excited state charge density, at the time of interest, with energy $\hbar\omega$ (at 400 fs in Figures 5(a) and 5(b) below). The rationale for doing so follows from the physics of lasers [43]: the probability of emission from the state with given energy is proportional to the number of excited atoms that have this energy.

### 3. Electronic Structure of bilayer MoS$_2$

In these single-layer TMDCs, the crystal structure is determined by just the lattice constant 'a'. Our PBE calculations for the lattice constants for the bulk, bilayer, and single-layer MoS$_2$ are 3.173 Å, 3.172 Å, and 3.170 Å, respectively. Using these optimized lattice constants, we have calculated the electronic band structure of bulk, bilayer, and single-layer MoS$_2$, along the lines connecting high-symmetry points of the Brillouin zone (BZ), which are displayed in Figure SI.1. The calculated band structure of all three systems are in good agreement with previous reports[44-47]. Our calculated electronic density of states (DOS) and the projected density of states (PDOS) on Mo-*d* and S-*p* orbitals for these three systems presented in Figure SI.2, show that the top of valence band and bottom of conduction band are contributed mainly by Mo-*d* orbitals, with a strong hybridization of Mo-*d* and S-*p* orbitals. The valence band tail in PDOS of bulk systems diminishes in case of bilayer and ultimately vanished in case of the single layer. This is because the valence band edge at Γ moves toward lower energy in case of bilayer and single layer, compared to that in the bulk, as we can see in Figure SI.1. The band structures in Figure SI.1 also verify that bulk and bilayer MoS$_2$ are indirect band-gap semiconductors with the valence band maximum (VBMAX) and conduction band minimum (CBMIN) located at the Γ and Q points of the BZ, respectively. We also note that it is evident in Figure SI.1 that the conduction band valley at the Q point starts shifting toward higher energy and the valence band hill at Γ point shifts toward lower energy, while the valence band hill at the K point shifts toward higher energy, as the thickness of the model system changes from bulk to two layers. Note that for the single layer, VBMAX and CBMIN are both located at the K point. From our calculations we find that the direct K→K gap for single-layer MoS$_2$ is 1.66 eV and the indirect Γ→ Q gaps for bulk and bilayer MoS$_2$ are 0.92 and 1.49 eV, respectively.

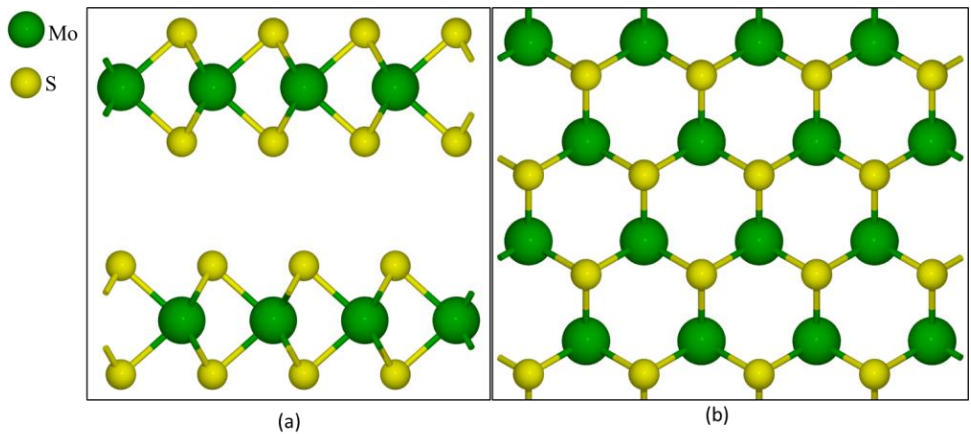

Figure 1: Schematic representation of bilayer MoS$_2$: (a) side view and (b) top view.

## 4. Electron-Phonon Coupling Coefficients for Bilayer MoS$_2$

We used the relaxed structure of bilayer MoS$_2$ presented above to calculate phonon dispersion curves for bilayer MoS$_2$ which are plotted in Figure SI.3(a). The features seen in Figure SI.3(a) are in general agreement with previous reports [30, 48, 49] and hence not presented here. In bilayer MoS$_2$, there are three acoustic branches: one longitudinal acoustic (LA), one transverse acoustic (TA), and one flexural acoustic (ZA). The LA mode reaches a value of 235 cm$^{-1}$ at M point, and 245 cm$^{-1}$ at K point. We note that there are no degeneracies at the M and K points, and the two crossings of the LA and TA branches just before and after the M point. The high-frequency optical modes are separated from the low-frequency modes by a gap of 41 cm$^{-1}$. We have drawn two horizontal blue dashed lines in the dispersion curve (Figure SI.4(a)) to depict the gap. In Figure SI.4(a) the in-plane and out of plane vibrational modes are represented by $E_{2g}$ and $A_{1g}$.

With $\omega_{qv}$ extracted from the calculated phonon dispersion curves (see Figure SI.4(a)) and using Eqs. (1 - 3), we calculate the phonon band index-dependent (phonon) linewidths and electron-phonon coupling coefficients which are summarized in Figure 2. The largest contribution to the linewidth comes from the $A_{1g}$ mode at the Γ point. As for the electron-phonon coupling, it is the LA mode at the K point that is most dominant, followed by the contribution from $A_{1g}$ mode at the Γ point. From the results in Figure 2(a), we can also infer that the lifetimes of the contributing phonon modes are longer than 16 ps, which means that their effect would be discernible in our analysis of the charge dynamics of the system at the hundreds of femtosecond time scale. Our results for the total and band-resolved $\alpha^2 F(\omega)$ (Eliashberg function) shown in Figure 3(a) and (b) are reflective of similar trends in the phonon DOS (see Figure SI.3(b)), demonstrating that phonons with energies 350-400 cm$^{-1}$ (43-50 meV) play a dominant dole in the dynamics of the system.

As an aside, we should mention that there are interesting differences in the calculated phonon linewidths (Figure SI.4(a)) electron-phonon coupling constants (Figure SI.4(b)), and the Eliashberg function (Figure SI.5) of single-layer $MoS_2$ from those of the bilayer presented in Figures. 2 and 3. In particular, in the case of the single-layer, the K- and Γ-point phonon modes are not dramatically different from the modes at other k-point modes, in contrast to the case of the bilayer. As we show below, a strong electron-phonon coupling at K is responsible for a strong indirect emission in the bilayer system. Comparison of the Eliashberg functions for the two systems shows that while in the bilayer $A_{2g}$ phonons (frequency ~400cm$^{-1}$) play a prominent role in the coupled electron-phonon system, in the monolayer the TA mode (~180cm$^{-1}$) is the dominant phonon band. On the other hand, the combined contribution of the $ZO_1$ and $ZO_2$ modes to the Eliashberg function in the single-layer system gives the second sharp peak at ~400cm$^{-1}$, similar to what is found here for the bilayer.

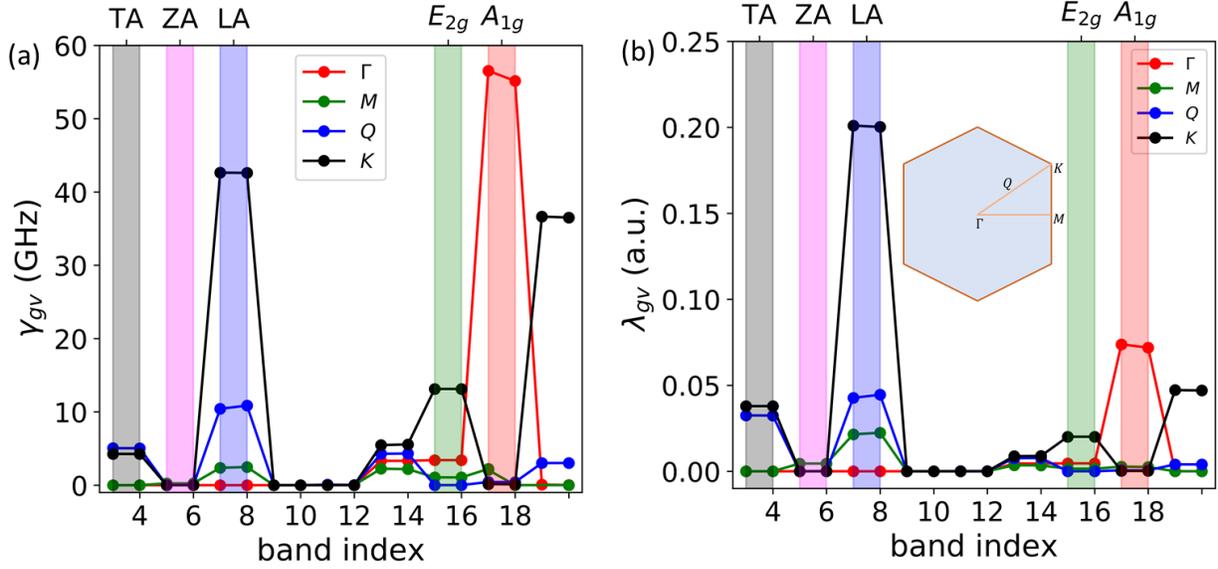

Figure 2: Calculated band index dependent phonon linewidths (a) and electron-phonon coupling constant (b) for the phonon bands at the special k-points of the two dimensional Brillioun zone (insets), for bilayer MoS$_2$.

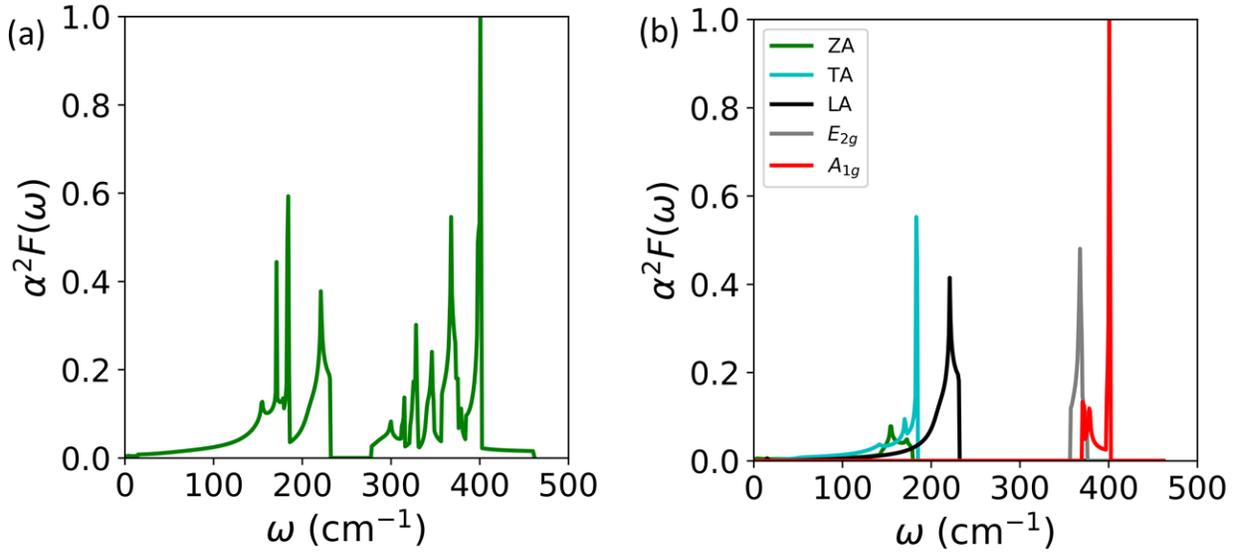

Figure 3: Eliashberg spectral function ($\alpha^2 F(\omega)$) for bilayer MoS$_2$: (a) contribution from all modes; (b) modes resolved contributions.

## 5. Electron-Phonon Mediated Relaxation of Excitation in Bilayer MoS$_2$

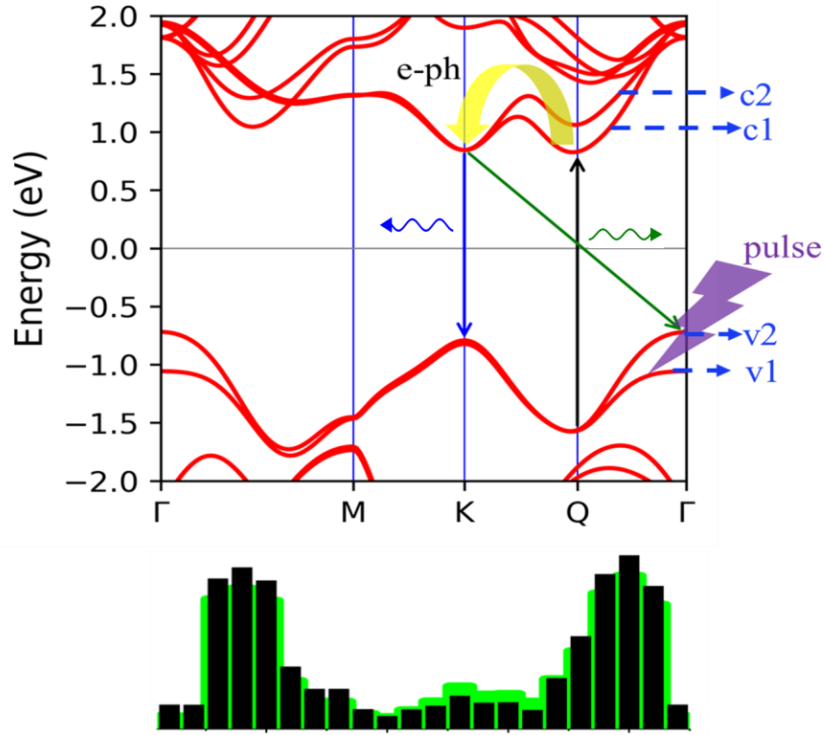

Figure 4: Top: Schematic representation of excitation and recombination processes that lead to photoluminescence. Bottom: relative excited charge accumulated, at k-points along the Brillouin zone, immediately following the external pulse (black) and after dynamics sets in (green).

To get physical insights into the excitation and emission process of the system at hand, we first calculated the momentum-resolved density of the pumped charges excited from the two top valence bands (labeled v1 and v2 in Figure 4) to the two lowest conduction bands (labeled c1 and c2 in Figure 4) in the presence of a 100 fs pulse of an electric field of magnitude 1V/Å, using Eq. (5) for the set of k-points shown in Figure 4 but with *no* electron-phonon interaction. We then calculated the excited charge density for the same set of k-points about 400 fs, during which electron-phonon scattering effects come into play (see Figure 5 and the discussion below). The results for both charge densities are shown in the bottom of Figure 4. The pulse excites electrons from the valence to the conduction bands through vertical transitions (photon momenta are very small). As one can see from Figure 4 (black columns) that the "pumped" excited charge is maximum between Q and Γ point of the BZ, which can be explained by a combination of two factors: high valence-bands DOS along these k-points (the bands are almost flat) and a relatively smaller separation, between the valence and conduction bands, compared to that at other k-points. The high density of the excited charges in another part of the BZ – between $\Gamma$ and M points – can be explained similarly. The results for the time dependence of the excited charge density at different k-points that correspond to the total pumped charges in Figure 4 are plotted in Figure 5(a). We normalized the charge densities at all k-points to the same saturated (after 200 fs) value for better visualization of their difference, once electron-phonon interaction is taken into account.

As shown in Figure 5(b), the occupancies at most of the k-points start to decrease after 200 fs, as electron-phonon coupling starts to play a role in scattering electrons out of the Q valley into the K valley. We can see that the originally highest-populated Q-valley gets significantly depopulated as a result of electron-phonon interaction, in contrast to what happens in the K-valley. The "final" excited charge density at different k-points is shown in the bottom of Figure 4(green columns). This electron relaxation due to electron-phonon interaction is represented schematically in the top panel of Figure 4. The energy required to move an electron from Q valley to K valley of conduction band can be estimated by the energy difference of these two valleys, ~ 40 meV. Recall from the analysis of the phonon spectrum that the effective (average) phonon energy is also about 40 meV (see Figure. SI.3(b) for the phonon DOS). Indeed phonons that populate this energy range must be responsible for the inter-valley charge transfer. We thus conclude that electron-phonon coupling is the main source of depopulation of excited state occupancies at points other than the K valley. At the K point, the major contribution to electron-phonon coupling arises from the low frequency longitudinal acoustic modes (see Figure 2(b)) which do not affect the occupancy of the excited state.

It is important to note that an inter-valley electron and hole dynamics was reported in work [27] for five-layer $MoS_2$. Similar to results obtained here, the inter-valley transitions occur at the few hundred-fs timescale. However, as a result of differences in the band order, transition take place for the five-layer system from K- to Q-valley (electrons) and from K- to $\Gamma$-valley (holes).

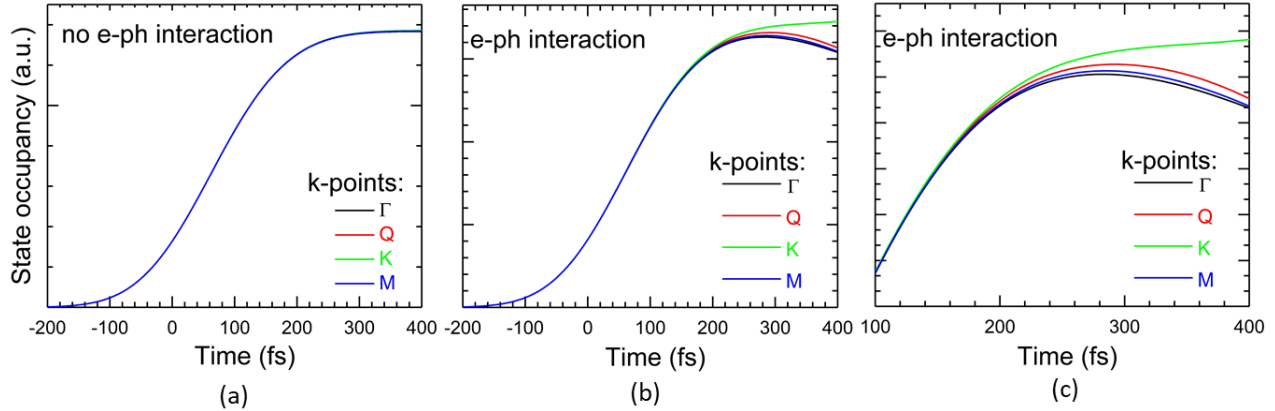

Figure 5: The time dependent occupancies of conduction band at special k-points ($\Gamma$, Q, K and M) of Brillouin zone. The state occupancies are plotted for the cases: (a) without electron-phonon interaction and (b) with electron-phonon interactions. In Fig (c) zoomed long-time occupancies from (b) are shown.

## 6. Calculated Emission Spectrum of Bilayer MoS$_2$

The emission spectra calculated without and with the electron-phonon dynamics are shown in Figure 6. We can see that the spectrum obtained without the electron-phonon scattering-induced charge redistribution (Figure 6(a)) has one dominating peak related to indirect emission attributed to the $K \to \Gamma$ transitions. The direct K-valley emission is reflected in a relatively weak shoulder in the spectrum. On the other hand, with electron-phonon induced redistributed charges we obtain a two-peak emission spectrum, as shown in Figure 6(b). The lower energy peak corresponds to indirect $K \to \Gamma$ transition while the peak toward higher energy corresponds to the direct K-valley transition. Transformation of the shoulder into a pronounced peak in emission is related to an enhanced charge density in the K-valley due to electron-phonon scattering (bottom Figure 4). It follows from Figure 2(b) that the strongest contribution to electron-phonon scattering comes from the LA phonon bands with momentum K. Thus, the indirect emission peak is formed mostly by electron transitions $K \to \Gamma$, i.e. from the K-valley in the conduction band that has an extra charge due to phonon-assisted charge transfer from Q to the K valley. In interpreting the emission spectra we note that electron-phonon scattering effects give rather good agreement with experiment regarding the position of the emission peaks found in Ref.[4] (Figure. 3) and Ref.[14] (Figure. 6). An extra small peak found in Ref. [4] may come from higher-energy exciton recombination. The difference in the relative magnitude of the peaks in Figure 6(b) are probably the result of the approximation that the probability of the emission transition is proportional to the amount of the excited charge density in the given valley (we take the charge density at maximum time used in the calculations (400 fs, Figure 5), which is still a transient time, i.e. the population of the valley can still change during total electron-phonon equilibration. It is important to stress that the emission peaks in this work come from the inter-band recombination of free electrons and holes, not exciton recombination. However, since the exciton binding energies for the different valleys are smaller than the corresponding bandgaps, the two-peak emission spectrum in Figure 6(b) will not change dramatically when the emission is caused by exciton recombination.

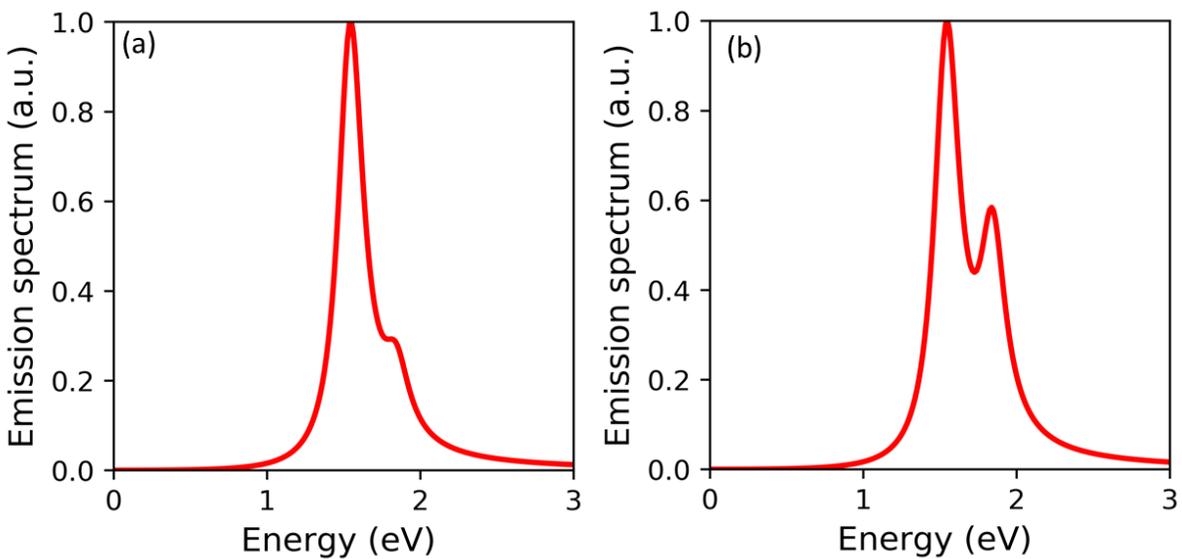

Figure 6: Emission spectrum of 2L MoS$_2$ calculated without (a) and with electron-phonon interactions included (b). The lower energy peak in emission corresponds to indirect emission, while the higher energy peak corresponds to direct emission.

## 7. Conclusions

Our investigation of the ultrafast excited charge dynamics and emission spectrum of bilayer MoS$_2$ demonstrate that electron-phonon interaction leads to transfer of electrons from the lowest-energy non-direct-gap Q valleys to the direct-gap K-valleys. This results in photoluminescence spectrum with peaks that correspond to the direct (K valley) and non-direct (LA phonon-assisted K to Γ valley) transitions. The emission spectrum obtained without the inclusion of intervalley dynamics consists of a single peak that corresponds to an indirect emission. The appearance of the second peak with the inclusion of electron-phonon interaction, demonstrates that the latter plays an important role in the ultrafast charge dynamics and photoluminescence in 2L MoS$_2$. The next step is to establish the contribution of the electron-electron and electron-hole interaction, including exciton effects, into the ultrafast dynamics and emission of the system. Work in this direction is in progress.

Acknowledgements

This work was supported by the United States Department of Energy under grant No. DE-FG02-07ER46354. The authors acknowledge the National Energy Research Scientific Computing Center for providing the computational resources (NERSC, project m3612). We are thankful to Lyman Baker for critical reading of the manuscript and to Duy Le and Peter Dowben for many useful discussions.

**Supplementary Information:**

## SI.1 The Liouville equations and the electron-phonon scattering terms

The density matrix satisfies the Liouville equation:

$$i\frac{\partial \rho_k^{lm}(t)}{\partial t} = [H,\rho]^{lm}(t) \equiv \sum_n \left(H_k^{ln}(t)\rho_k^{nm}(t) - \rho_k^{ln}(t)H_k^{nm}(t)\right) + \left(\frac{\partial \rho_k^{lm}}{\partial t}\right)_{scatt}, \qquad \text{(SI.1)}$$

where

$$H_k^{ml}(t) = \int \psi_k^{m*}(r)\hat{H}(r,t)\psi_k^l(r)dr. \qquad \text{(SI.2)}$$

The Hamiltonian in the integral in Eq. (SI.2)

$$\hat{H}(r,t) = -\frac{\nabla^2}{2m} + V_{ion}(r) + V_H[n](r) + V_{XC}[n](r) + e\vec{r}\vec{E}(t). \qquad \text{(SI.3)}$$

consists of the DFT Kohn-Sham Hamiltonian (the first four terms) and the laser-pulse perturbation potential (the last term). The time-dependence of the applied Gaussian pulse is

$$\vec{E}(t) = \vec{E}_0 e^{-\frac{t^2}{\tau^2}}, \qquad \text{(SI.4)}$$

where $\vec{E}_0 = E_0(1,1,1)$ and $E_0 = 1V/A$ and $\tau = 100fs$ is the pulse duration.

Since

$$\left[-\frac{\nabla^2}{2m} + V_{ion}(r) + V_H[n](r) + V_{XC}[n](r)\right]\psi_k^m(r) = \varepsilon_k^m \psi_k^m(r), \qquad \text{(SI.5)}$$

the matrix elements (SI.2) have a simple form

$$H_k^{ml} = \varepsilon_k^m \delta^{ml} + \vec{d}_k^{ml}\vec{E}(t), \qquad \text{(SI.6)}$$

where

$$\vec{d}_k^{ml} = e\int \psi_k^{m*}(r)\vec{r}\psi_k^l(r)dr, \qquad \text{(SI.7)}$$

are the transition dipole moments.

Finally, substituting Eq.(SI.6) into Eq.(SI.1) one obtains the explicit form of the Liouville equations:

$$i\frac{\partial \rho_k^{lm}(t)}{\partial t} = (\varepsilon_k^l - \varepsilon_k^m)\rho_k^{lm}(t) + \vec{E}(t)\sum_n \left(\vec{d}_k^{ln}\rho_k^{nm}(t) - \rho_k^{ln}(t)\vec{d}_k^{nm}(t)\right) + \left(\frac{\partial \rho_k^{lm}}{\partial t}\right)_{scatt}. \qquad \text{(SI.8)}$$

The diagonal and non-diagonal scattering matrix elements in Eq. (SI.8) have the following form:

$$\left(\frac{\partial \rho_k^{ll}}{\partial t}\right)_{scatt} = -2\pi \sum_{q,\nu} g_{q\nu}^2 \delta(\varepsilon_{k+q}^l - \varepsilon_k^l - \omega_0)\{N(\omega_0)\rho_k^{ll}(t)[1 - \rho_{k+q}^{ll}(t)]$$
$$- (N(\omega_0) + 1)\rho_{k+q}^{ll}(t)[1 - \rho_k^{ll}(t)]\}$$
$$-2\pi \sum_{q,\nu} g_{q\nu}^2 \delta(\varepsilon_{k-q}^l - \varepsilon_k^l + \omega_0)\{(N(\omega_0) + 1)\rho_k^{ll}(t)[1 - \rho_{k-q}^{ll}(t)]$$
$$- N(\omega_0)\rho_{k-q}^{ll}(t)[1 - \rho_k^{ll}(t)]\}, \quad (SI.9)$$

$$\left(\frac{\partial \rho_k^{lm}}{\partial t}\right)_{scatt}$$
$$= -2\pi \sum_{q,\nu,\sigma=\pm 1} g_{q\nu}^2 \delta(\varepsilon_{k+q}^l - \varepsilon_k^m - \sigma\omega_0)\{N(\sigma\omega_0)[\rho_k^{lm}(t)\rho_{k+\sigma q}^{mm}(t) - \rho_{k+\sigma q}^{lm}(t)\rho_k^{ll}(t)]$$
$$+ [N(\sigma\omega_0) + 1][\rho_k^{lm}(t)\rho_{k+\sigma q}^{ll}(t) - \rho_{k+\sigma q}^{lm}(t)\rho_k^{mm}(t)]\} - \{k \leftrightarrow k + \sigma q\}, (SI.10)$$

where $N(\omega_0) = 1/[\exp[\omega_0/T] - 1]$ is the Bose distribution function with used temperature $T = 0.1 eV$ and $\omega_0 = 40 meV$ is the effective phonon frequency.

**SI.2: Band Structure, Electronic Density of States and Phonon Dispersion of Bilayer MoS$_2$**

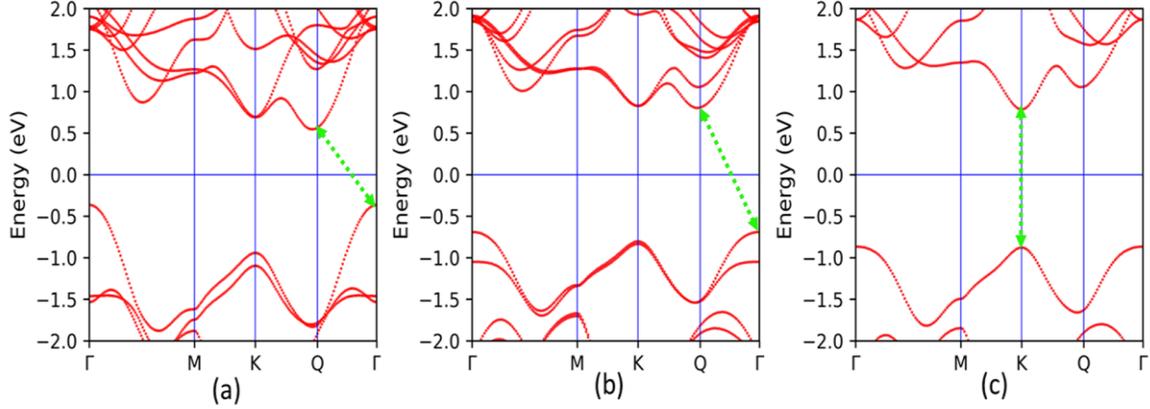

Figure SI.1: Band structure of (a) bulk (b) bilayer (c) and single-layer MoS$_2$. Fermi level (horizontal blue line) is set to zero. The green dotted line shows the indirect gap for both the bulk and bilayer, and direct band gap for single-layer cases.

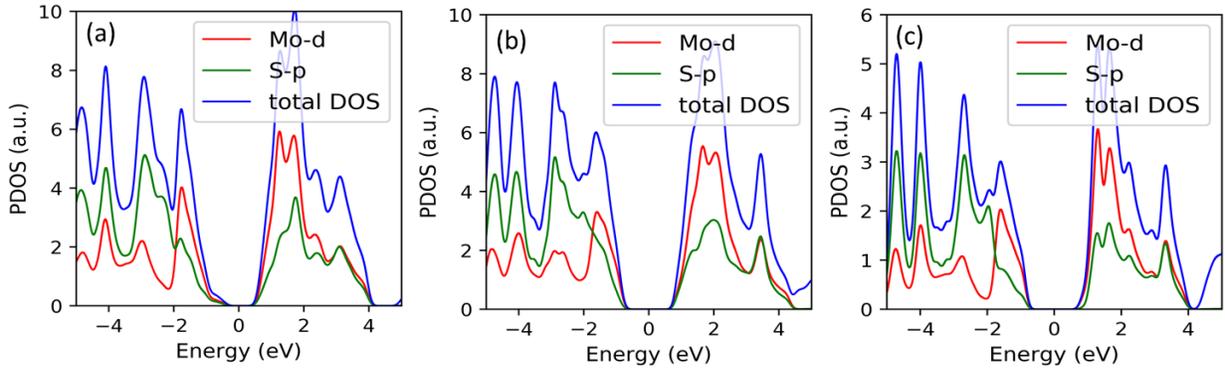

Figure SI.2: Total and Projected Density of States of (a) bulk, (b) bilayer and (c) single-layer $MoS_2$. The blue lines represent total density of states; green and red lines represent the projected density of states of S-p and Mo-d orbitals, respectively

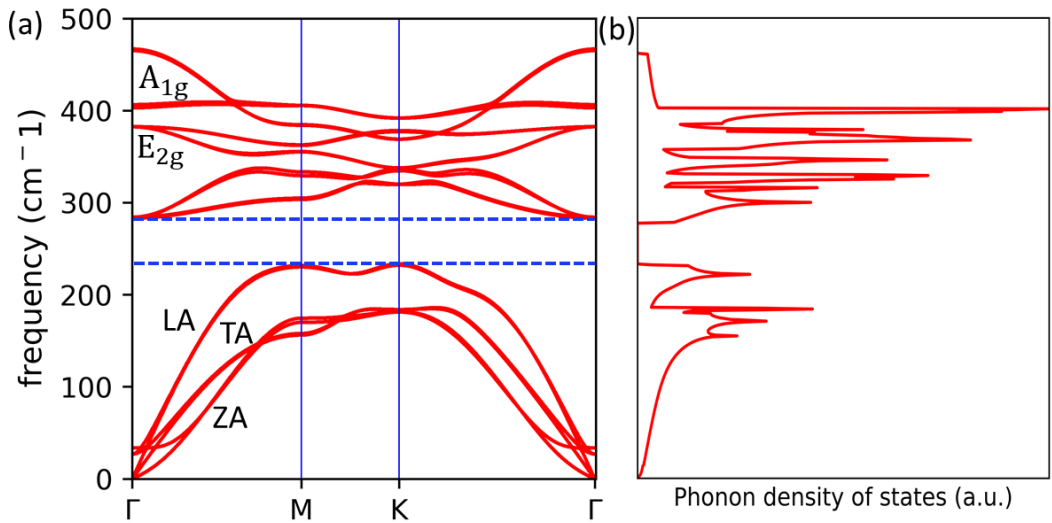

Figure SI.3: Phonon dispersion for bilayer $MoS_2$ (a) and phonon density of states (b) The horizontal dashed lines in (a) show the separation/gap of low frequency and high frequency phonon branches.

**SI.3: Electron Phonon Coupling, Linewidths and Eliashberg function Single layer $MoS_2$**

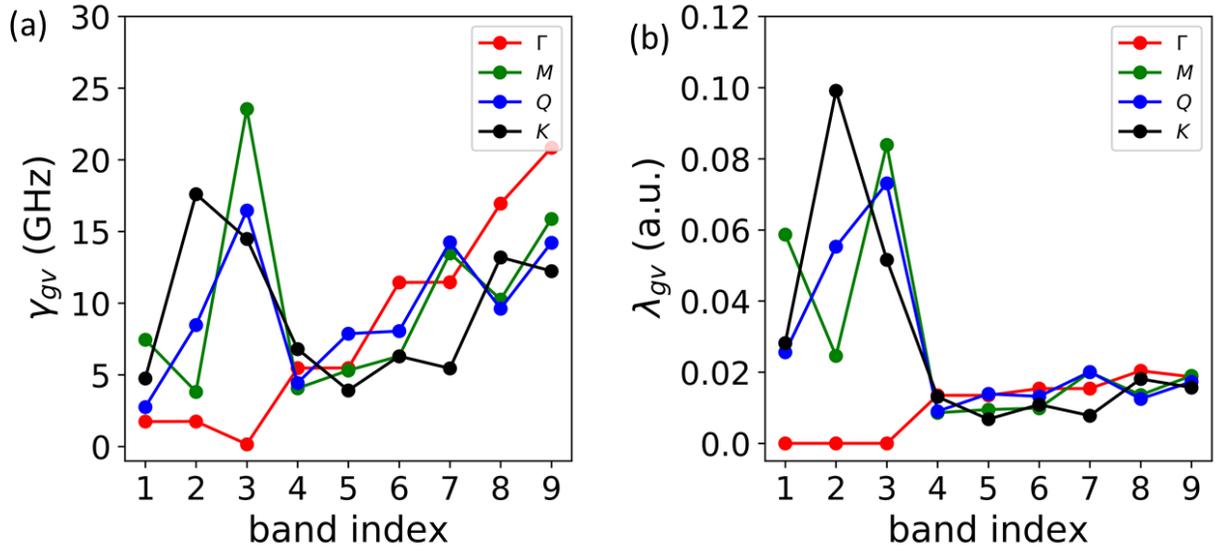

Figure SI.4: Calculated band index dependent phonon linewidths (a) and electron-phonon coupling constant (b) for the phonon bands at the special k-points of the two dimensional Brillioun zone, for single layer $MoS_2$.

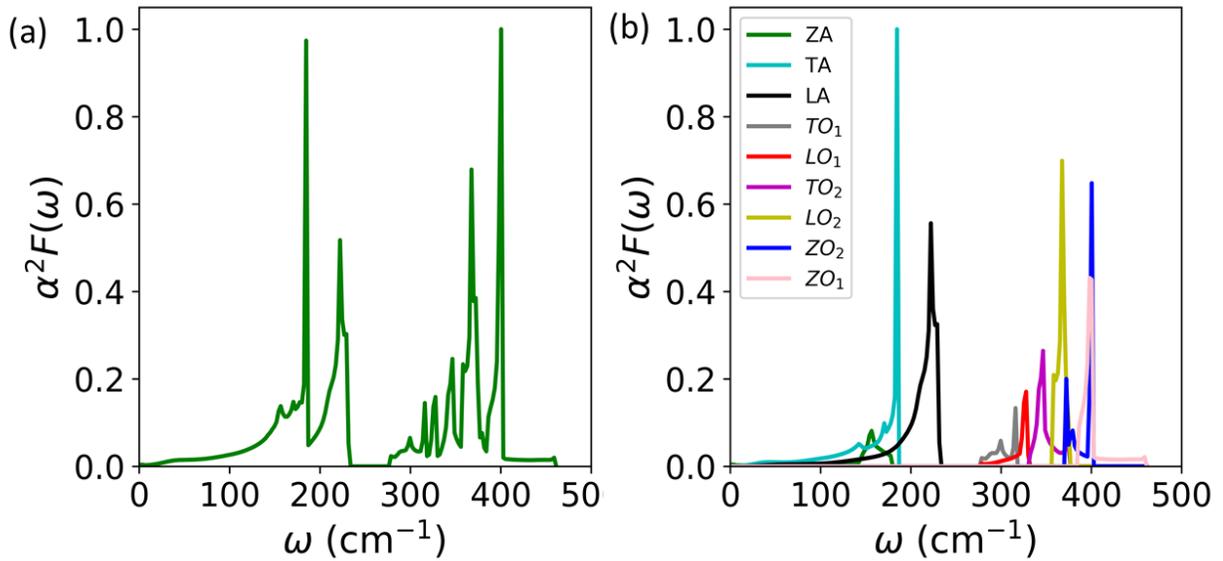

Figure SI.5: Eliashberg spectral function ($\alpha^2F(\omega)$) for single layer $MoS_2$: (a) contribution from all modes; (b) modes resolved contributions